\documentclass[preprintnumbers,superscriptaddress,amsmath,amssymb,preprint,nofootinbib]{revtex4}
\usepackage{graphicx}
\usepackage{bbm}
\usepackage{bm}
\usepackage{amsfonts}
\usepackage[colorlinks,linkcolor=red,anchorcolor=blue,citecolor=green]{hyperref}
\usepackage{subfigure}
\usepackage{float}
\usepackage{enumerate}
\newcommand\diff{\mathrm{d}}

\newcommand\e{\mathrm{e}}

\begin{document}
\title{
	Probing the conformal invariance around the nonsingular static spherical black holes with waves
}
\author{Rui-Hui \surname{Lin}}
\email[]{linrh@shnu.edu.cn}
\author{Rui \surname{Jiang}}
\affiliation{Division of Mathematical and Theoretical Physics, Shanghai Normal University,
100 Guilin Road, Shanghai 200234, China}
\author{Yang \surname{Huang}}
\email[]{sps_huangy@ujn.edu.cn}
\affiliation{Division of Mathematical and Theoretical Physics, Shanghai Normal University,
100 Guilin Road, Shanghai 200234, China}
\affiliation{School of Physics and Technology, University of Jinan,
No. 336, West Road of Nan Xinzhuang, Jinan 250022, Shandong, China}
\author{Xiang-Hua \surname{Zhai}}
\email[]{zhaixh@shnu.edu.cn}
\affiliation{Division of Mathematical and Theoretical Physics, Shanghai Normal University,
100 Guilin Road, Shanghai 200234, China}
\begin{abstract}
	Conformal invariance can ameliorate or eliminate the singularities residing in the black holes,
	and may still exist in the strong gravity regimes close to these black holes.
	In this paper, we try to probe this conformal invariance
	by looking into the wave absorption and scattering by the nonsingular static spherical black holes.
	The partial and total absorption cross section,
	as well as the differential scattering cross section,
	are presented for black holes with different choices of conformal parameters.
	Although the photon trajectories are unchanged from the Schwarzschild case since the spacetimes are conformally related,
	the wave optics are affected by the conformal parameters.
	As a result, the absorption of waves generally increases with the conformal parameters,
	while the shadow of the black holes remains the same as the Schwarzschild case.
	Moreover, the peaks in the oscillatory pattern of scattering shift towards smaller observing angles as the conformal parameters grows,
	while the widths of the glory peaks do not show sensitive dependence.
	The unique signature of the wave absorption and scattering by the nonsingular static spherical black holes in conformal gravity
	thus can serve to distinguish themselves from the Schwarzschild in the low frequency regime,
	and from other spherical black holes of alternative gravities in the high frequency limit and glory peaks.
\end{abstract}

\maketitle
\section{Introduction}
As a great theoretical achievement,
Einstein's General Relativity (GR) works perfectly well in explaining various observations.
A class of its solutions, black holes, also seems to prove present in binary systems and galactic centers
by the growing evidence in recent years\cite{PhysRevLett.116.061102,PhysRevLett.119.141101,Akiyama:2019cqa}.
Nonetheless, the long known challenges of GR,
such as the dark contents of the universe and
the quantization of the gravity,
have motivated the search of alternative theories of gravity.
The detection of gravitational waves (GWs)\cite{PhysRevLett.116.061102,PhysRevLett.119.141101}
and the direct observation of the center of our Galaxy by Event Horizon Telescope\cite{Akiyama:2019cqa}
may thus herald a new era of testing those alternative gravity theories
that have different predictions from GR in strong field regime.
One of the schemes is the conformal gravity
which can ameliorate or eliminate the singularities in black hole solutions of GR\cite{ENGLERT1976407,Naelikar1977}.
This description of gravity is invariant under a conformal transformation of the metric
\begin{equation}
	g_{\mu\nu}\rightarrow \hat{g}_{\mu\nu}=S(x)g_{\mu\nu},
	\label{conformaltransf}
\end{equation}
where $S(x)$ is a nonsingular function depending on the spacetime coordinates.
Conformal gravity can be constructed with, for example, the Weyl tensor
\cite{ENGLERT1976407,Naelikar1977,PhysRevLett.53.315,Mannheim2012,PhysRevD.94.086008,MANNHEIM2017125}
or the dilaton\cite{Darowski2009,tHooft2011,Bambi_2017}.
With this conformal invariance, the singularities of spacetime can be avoided
in the sense that they are artificial and rely on the choice of conformal gauges\cite{Naelikar1977,Bambi_2017},
similar to the situations of the apparent or coordinate singularities in GR.

However, the conformal invariance is not really possessed by the realistic spacetime\cite{Bambi_2017}.
For conformal gravity to be a physical description of gravitation,
some sort of symmetry breaking mechanism is thus required,
such that the spacetime can present itself in the current phase.
In fact, if we consider specifically the conformal gravity constructed with the dilaton $\varphi$\cite{Bambi_2017,tHooft2011,Darowski2009},
\begin{equation}
	\mathcal S=\int\diff^4x \sqrt{-\hat{g}} \left( \varphi^2\hat{R}+6\hat{g}^{\mu\nu}\partial_\mu\varphi\partial_\nu\varphi \right),
	\label{dilatonaction}
\end{equation}
which is invariant under the transformation Eq.\eqref{conformaltransf} with $\varphi\rightarrow S^{-1}\varphi$,
the Einstein-Hilbert Lagrangian can be recovered as one of the possible vacua
in the spontaneous breaking of the conformal symmetry in analogy with the Higgs mechanism\cite{Bambi_2017}.
And in the strong field regime such as the regions close to the black holes,
the spacetime may be still in its conformally invariant phase\cite{RevModPhys.54.729,tHooft2011,Mannheim2012}.
It was shown that a class of nonsingular black holes from conformal gravity
is compatible with the x-ray observation data\cite{PhysRevD.95.064006,PhysRevD.98.024007}.
Various aspects of this class of black holes including the formation\cite{Bambi2018},
evaporation\cite{Bambi_2017eva} and perturbation\cite{PhysRevD.96.064028,PhysRevD.99.104003} have also been discussed.

In this work,
we adopt another approach, namely the wave absorption and scattering, to study this conformal invariance in the region around the nonsingular black holes.
The study of black hole scattering dates back to 1960s\cite{Matzner1968,PhysRevD.18.1030,PhysRevD.18.1798,Futterman1988,PhysRevD.46.4477}.
Using waves as probes may provide deeper insights of the physics of black holes and high curvature spacetime.
And increasing amount of attention has been attracted to these subjects related to various gravity theories and black holes in recent years.
(See, e.g., Refs.\cite{Glampedakis_2001,Jung_2004,Dolan_2008,PhysRevD.79.064022,PhysRevLett.102.231103,Gonzalez2010,Gonzalez:2010ht,liao2012,Kanai_2013,Raffaelli:2013ih,PhysRevD.88.064033,PhysRevD.89.104053,PhysRevD.92.024012,Nambu:2015aea,LEITE2017130,deOliveira2018} and the references therein.)
Besides the theoretical interests,
the experimental advances, including the historical detection of GWs\cite{PhysRevLett.116.061102,PhysRevLett.119.141101} and
the first observation of the supermassive black hole in M87 by Event Horizon Telescope\cite{Akiyama:2019cqa},
have also incentivized the growing investigation of the interaction between black holes and various waves.
The numerical techniques for related calculation have also been significantly improved\cite{PhysRev.95.500,Konoplya_2019,OuldElHadj:2019kji}.
The pattern of the scattered waves can be seen far away from the black hole,
and may likely bear the information of the near-horizon structure of the black hole,
or, in the case of this work, the information of conformal invariance.

Following this path,
in this paper,
we try to tell apart the nonsingular static spherical black holes from the Schwarzschild and other spherical ones
by examining the absorption and scattering of scalar waves.
The paper is organized as follows.
In Sec. \ref{sec:bhs}, we briefly review the nonsingular black holes in the conformal gravity.
Section \ref{sec:absorption} and \ref{sec:scattering} are dedicated to the absorption and scattering cross sections of scalar waves, respectively.
We conclude in Sec.\ref{summary}.
Throughout the paper, we use the units that $c=8\pi G=1$,
and assume that the impinging waves have no backreaction on the background.

\section{Nonsingular static spherical black holes in conformal gravity}%
\label{sec:bhs}
In conformal gravity,
the spacetime singularity residing in the black hole can be considered as mathematical artificial and
can be removed through conformal transformation.
For static spherical case in GR, the Schwarzschild black hole is described by
\begin{equation}
	\diff\tau_\text{s}^2=f(r)\diff t^2-\frac{1}{f(r)}\diff r^2-r^2\diff\Omega^2,
	\label{schw}
\end{equation}
where
\begin{equation}
	f(r)=1-\frac{r_\text s}{r},
	\label{schf}
\end{equation}
and $r_\text s$ is the horizon radius.
In this case,
the intrinsic singularity at $r=0$ can be removed by considering a conformally related metric
\begin{equation}
	\diff\tau^2=\hat{g}_{\mu\nu}\diff x^\mu\diff x^\nu=S(r)\diff\tau_\text{s}^2
	\label{conformalschw}
\end{equation}
with a suitable conformal factor $S(r)$.
One particular choice of $S(r)$ is\cite{Bambi_2017}
\begin{equation}
	S(r)=\left( 1+\frac{L^2}{r^2} \right)^{2N},
	\label{Sr}
\end{equation}
with a positive integer $N$ and a new length scale $L$.
Obviously, the Schwarzschild metric is recovered when $L\ll r$.
It is shown that $L$ may be restricted to $L\le0.6r_\text s$ for related cases by observation\cite{PhysRevD.95.064006,PhysRevD.98.024007}.
In this work,
following the orders of magnitude in Refs.\cite{PhysRevD.96.064028},
we consider $N$ ranges up to $50$ and $L$ ranges up to $0.6r_\text s$.

The spacetime described by Eq.\eqref{conformalschw} is nonsingular everywhere \cite{Bambi_2017}.
Particularly, at $r=0$,
the Ricci scalar $\hat R$ can be approximated as
\begin{equation}
	\hat R\simeq\frac{24N^2r_\text{s}}{L^{4N}}r^{4N-3},
	\label{ricci}
\end{equation}
and the Krestchmann scalar $\hat{\mathcal K}=\hat{R}_{\alpha\beta\gamma\delta}\hat{R}^{\alpha\beta\gamma\delta}$ as
\begin{equation}
	\hat{\mathcal K}\simeq\frac{12\left( 1+12N^2-16N^3+16N^4 \right)r^2_\text s}{L^{8N}}r^{8N-6}.
	\label{krestchmann}
\end{equation}
Since it is presumed that $N$ is a positive integer,
these curvature invariants are thus regular everywhere.
It is also shown that
the spacetime described by Eq.\eqref{conformalschw} is geodesically complete
in that massless particles cannot reach the center for a finite value of the affine parameter
and massive particles cannot reach the center in a finite proper time\cite{Bambi_2017}.

To consider the classical paths
in static spherical spacetime conformally related to Schwarzchild metric,
one can read from the metric Eqs.\eqref{schw} and \eqref{conformalschw} that
\begin{equation}
	\dot\tau^2=S(r)\left[ f(r)\dot t^2-\frac{\dot r^2}{f(r)}-r^2\dot\theta^2-r^2\sin^2\theta\dot\phi^2 \right]=\kappa,
	\label{dots}
\end{equation}
where the overdot indicates the derivative with respect to certain affine parameter,
and $\kappa=0,1$ for massless and massive particles, respectively.
In the current case with $\kappa=0$,
although the conformal factor $S(r)$ can be canceled out,
it still affects the movement of the massless particles by altering the conserved energy and angular momentum (see, e.g., Ref.\cite{Bambi_2017}):
\begin{equation}
	e=S(r)f(r)\dot t,\quad j=S(r)r^2\dot\theta,
	\label{conservedel}
\end{equation}
where we have considered the plane with $\diff\phi=0$.
Nonetheless, the trajectories of massless particles are still unchanged,
and the null orbit equation can be written as
\begin{equation}
	\left( \frac{\diff u}{\diff\theta} \right)^2=\frac{1}{b^2}-u^2+r_\text su^3,
	\label{schwgeo}
\end{equation}
with $u\equiv1/r$,
where $b$ is the impact parameter, i.e., the ratio of conserved angular momentum and energy of the particle.
Hence the deflection function $b(\theta)$
can also be approximated by\cite{Darwin1959,*Darwin1961}
\begin{equation}
	\frac{b(\theta)}{r_\text s}\sim\frac{3}{2}\sqrt{3} +1.74\e^{-\theta},\quad\text{for}\quad\theta\gtrsim\pi.
	\label{darwin}
\end{equation}
The critical impact parameter,
which leads to the unstable orbiting of massless particles trapped on the photon sphere,
is thus also $b_\text s=\frac32\sqrt{3}r_\text s $.
The shadow area $A_\text{shadow}$ of the current black holes is therefore also $A_\text{shadow}=\frac{27}{4}\pi r^2_\text s$.
This suggests that,
if seen afar,
the conformal invariance does not bear much difference in geometric optics compared to the Schwarzschild spacetime.
However, since other spherical black holes, such as
the Reissner-Nordstr\"om black hole\cite{PhysRevD.79.064022},
the Bardeen black hole\cite{PhysRevD.92.024012},
the black holes in Ho\v rava-Lifshitz gravity\cite{liao2012}
and the brane-world models\cite{deOliveira2018},
yield different null trajectories depending on the model parameters,
the shadows of the nonsingular black holes can be distinguished from these black holes.
Moreover, as we will see in the following sections,
the nonsingular static spherical black holes in conformal gravity
will demonstrate distinct absorption and scattering cross sections.

\section{Absorption cross section}%
\label{sec:absorption}
To consider the interaction between massless scalar wave and the black hole described by Eq.\eqref{conformalschw},
we firstly consider the absorption of the wave by the black hole utilizing the partial wave method.

\begin{figure}[!htpb]
	\centering
	\includegraphics[width=0.75\linewidth]{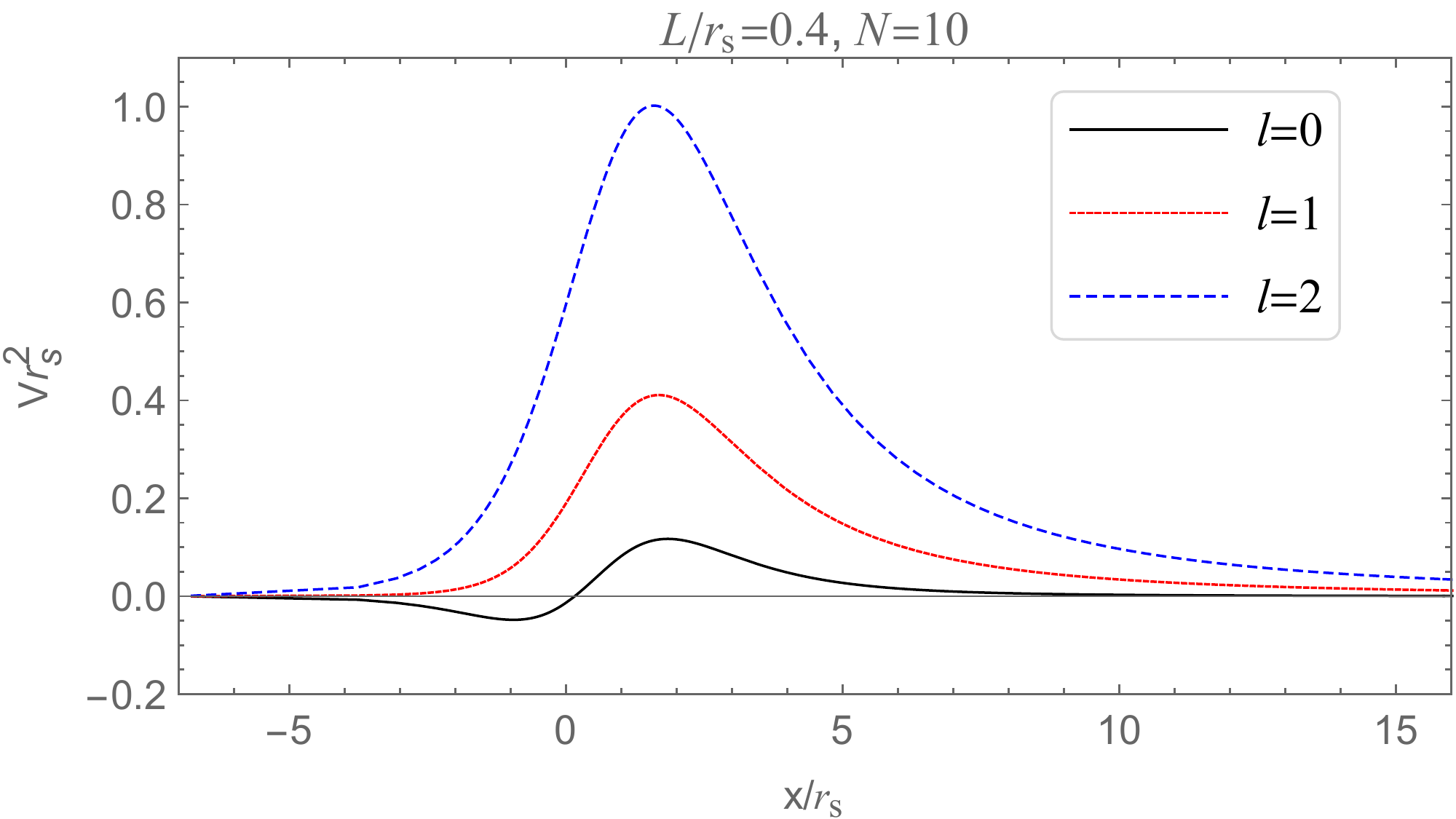}
	\includegraphics[width=0.75\linewidth]{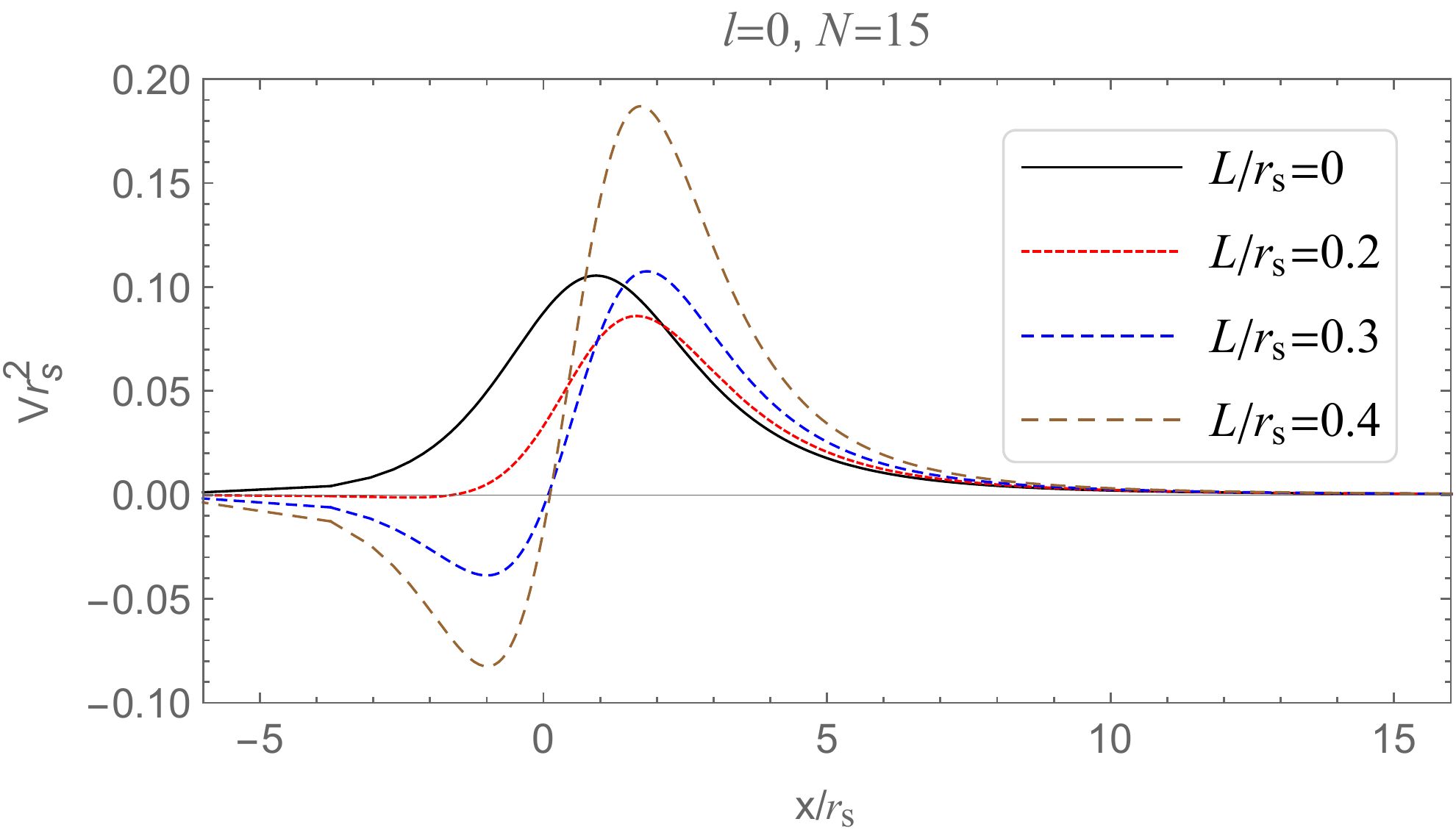}
	\includegraphics[width=0.75\linewidth]{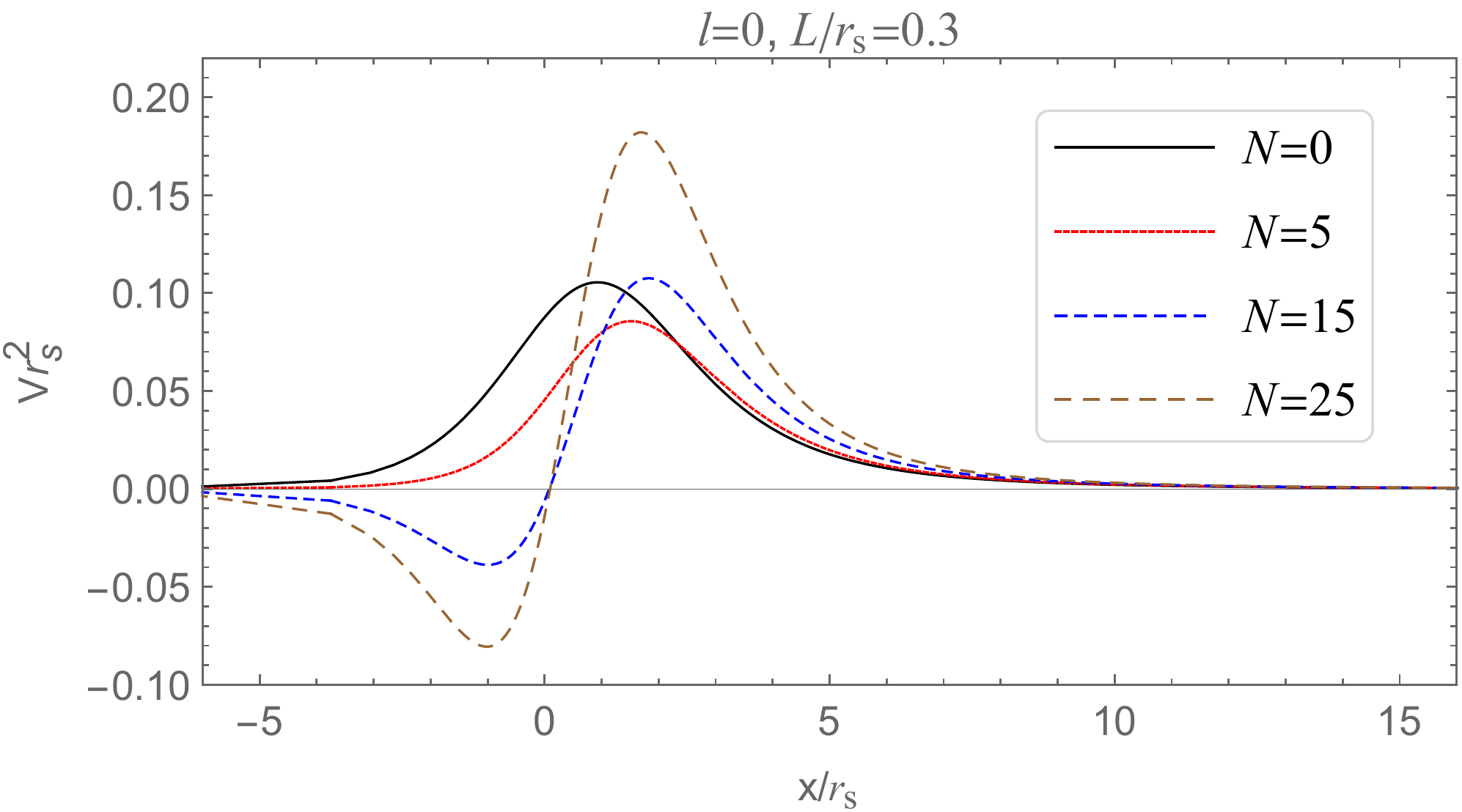}
	\caption{
		The dependence of the effective potential
		on the angular quantum number $l$ and
		the conformal parameters $L$ and $N$.
}
	\label{pot}
\end{figure}

A minimally coupled massless scalar field $\Phi$ in curved spacetime is described by the covariant Klein-Gordon equation
\begin{equation}
	\nabla^\mu\nabla_\mu\Phi=\frac{1}{\sqrt{-\hat g} }\partial_\mu\left( \hat g^{\mu\nu}\sqrt{-\hat g} \partial_\nu\Phi \right)=0.
	\label{kgeq}
\end{equation}
Separating the variables of $\Phi(t,r,\theta,\phi)$ as
\begin{equation}
	\Phi=\frac{\psi_{\omega l}(r)}{r\sqrt{S(r)} }Y_{lm}(\theta,\phi)\e^{-i\omega t},
	\label{seperate}
\end{equation}
where $Y_{lm}$ is the spherical harmonic function of degree $l$ and order $m$,
we have the equation for the radial function $\psi_{\omega l}(r)$
\begin{equation}
	\frac{\diff^2}{\diff x^2}\psi_{\omega l}(x)+\left[ \omega^2-V_\text{eff} \right]\psi_{\omega l}(x)=0,
	\label{kgradial}
\end{equation}
with the tortoise (Regge-Wheeler) coordinate $x$
\begin{equation}
	\frac{\diff}{\diff x}=f(r)\frac{\diff}{\diff r},
	\label{tortoise}
\end{equation}
and the effective potential related to the metric \eqref{conformalschw}
for the angular momentum index $l$
\begin{equation}
	V_\text{eff}\equiv f(r)\left\{ \frac{l(l+1)}{r^2}+\frac{1}{r\sqrt{S} }\frac{\diff}{\diff r}\left[ f(r)\frac{\diff}{\diff r}\left(r\sqrt{S} \right) \right] \right\}.
	\label{Veff}
\end{equation}
Some representative cases of the effective potential are plotted in Fig.\ref{pot}.
For $l=0$ part, which plays the major role in the low frequency absorption,
one can see that the potential barrier may be lower than the Schwarzschild case for small conformal parameters $L$ and $N$,
but as $L$ or $N$ gets larger,
the height of the potential barrier increases,
and will surpass the Schwarzschild barrier.
Moreover, on the side of the potential barrier close to the horizon,
there exists a potential well,
whose depth also increases with the conformal parameter $L$ and $N$.

Since the effective potential is localized and it vanishes at both the horizon ($x\rightarrow-\infty$) and
in the regime far away from the horizon ($x\rightarrow+\infty$),
we have the asymptotic solution to Eq.\eqref{kgradial}
\begin{equation}
	\psi_{\omega l}(x)\simeq
	\begin{cases}
		\mathcal T_{\omega l}\e^{-i\omega x}&\left( x\rightarrow-\infty \right),\\
		\e^{-i\omega x}+\mathcal R_{\omega l}\e^{i\omega x}&\left( x\rightarrow+\infty \right),
	\end{cases}
	\label{asymptpsi}
\end{equation}
where the amplitude of the impinging wave has been normalized,
and $\mathcal T_{\omega l}$ and $\mathcal R_{\omega l}$ are complex constants
satisfying $|\mathcal T_{\omega l}|^2+|\mathcal R_{\omega l}|^2=1$.
In the intermediate region $-\infty<x<+\infty$, one thus can match the two parts of Eq.\eqref{asymptpsi}
via Eq.\eqref{kgradial} numerically.

The partial and total absorption cross section are then given by
\begin{equation}
	\sigma_\text{abs}^{(l)}=\frac{\pi}{\omega^2}\left( 2l+1 \right)\left( 1-\left| \mathcal R_{\omega l} \right|^2 \right),
	\label{partialabs}
\end{equation}
and
\begin{equation}
	\sigma_\text{abs}=\sum_{l=0}^{\infty}\sigma_\text{abs}^{(l)},
	\label{totalabs}
\end{equation}
respectively.
\begin{figure}[!htpb]
	\centering
	\includegraphics[width=0.8\linewidth]{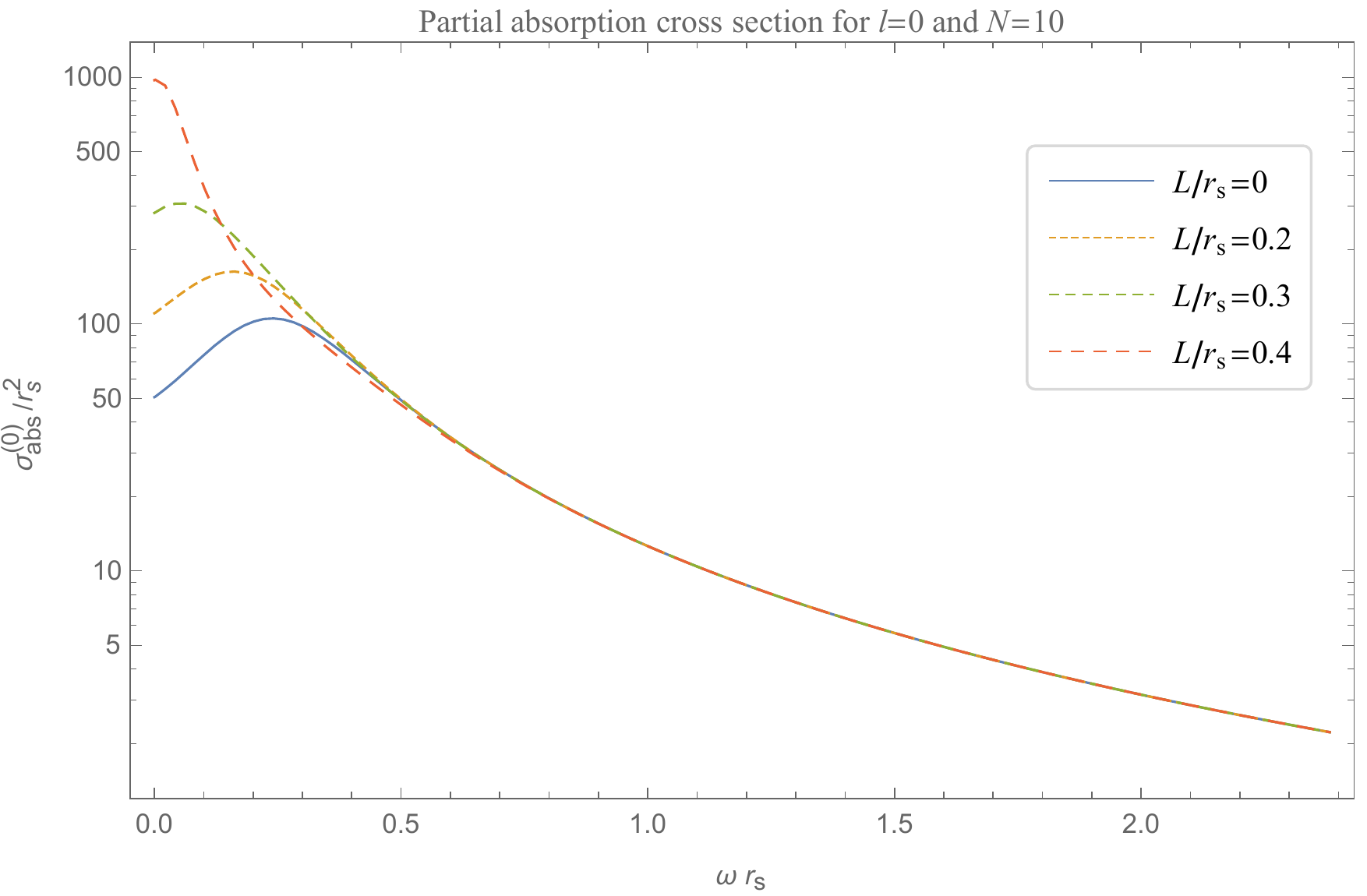}
	\includegraphics[width=0.8\linewidth]{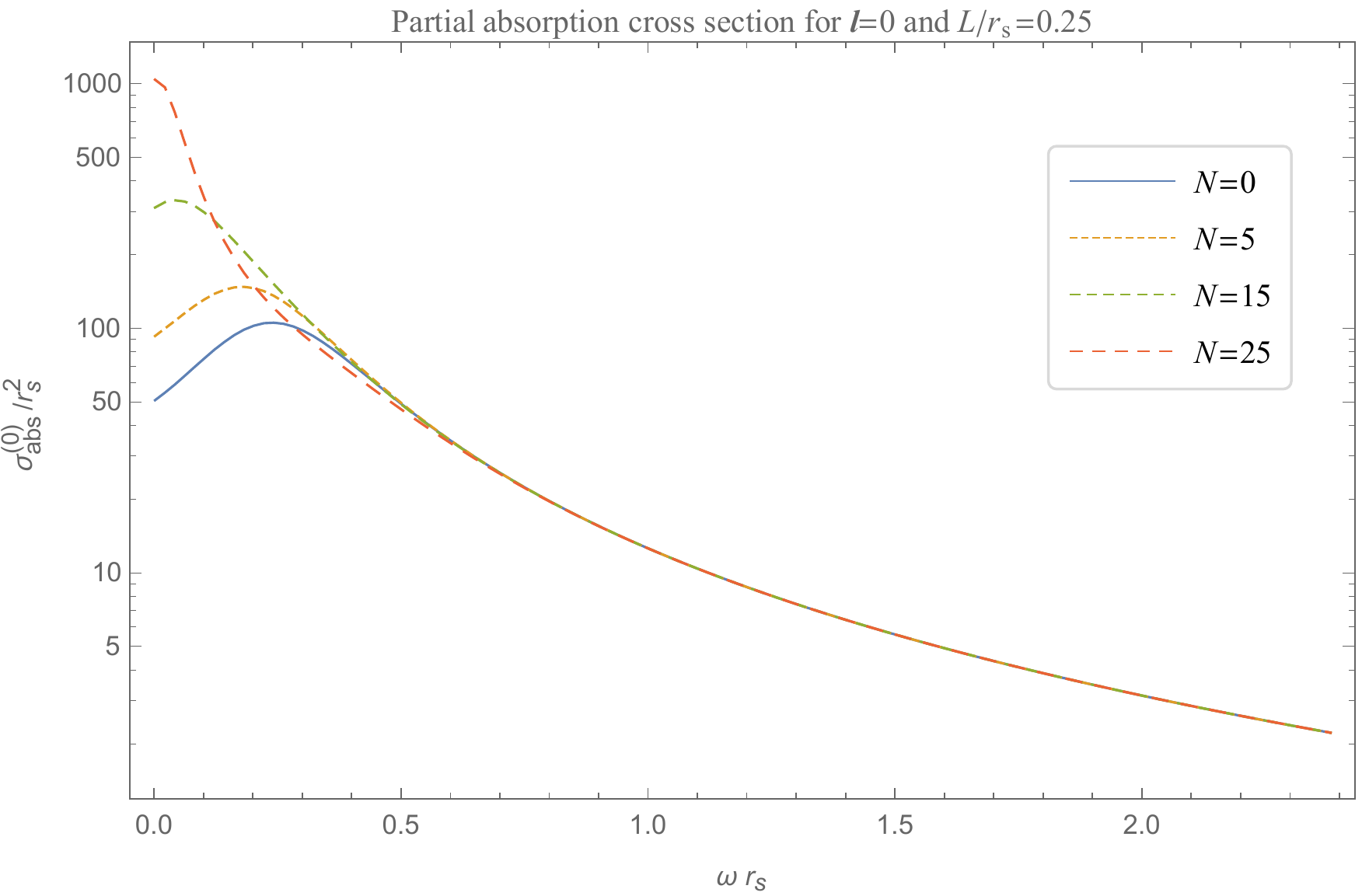}
	\caption{The partial absorption cross sections with $l=0$ and various $N$ or $L$
	for nonsingular static spherical black holes in conformal gravity.
}
	\label{absl=0}
\end{figure}
\begin{figure}[!htpb]
	\centering
	\includegraphics[width=0.8\linewidth]{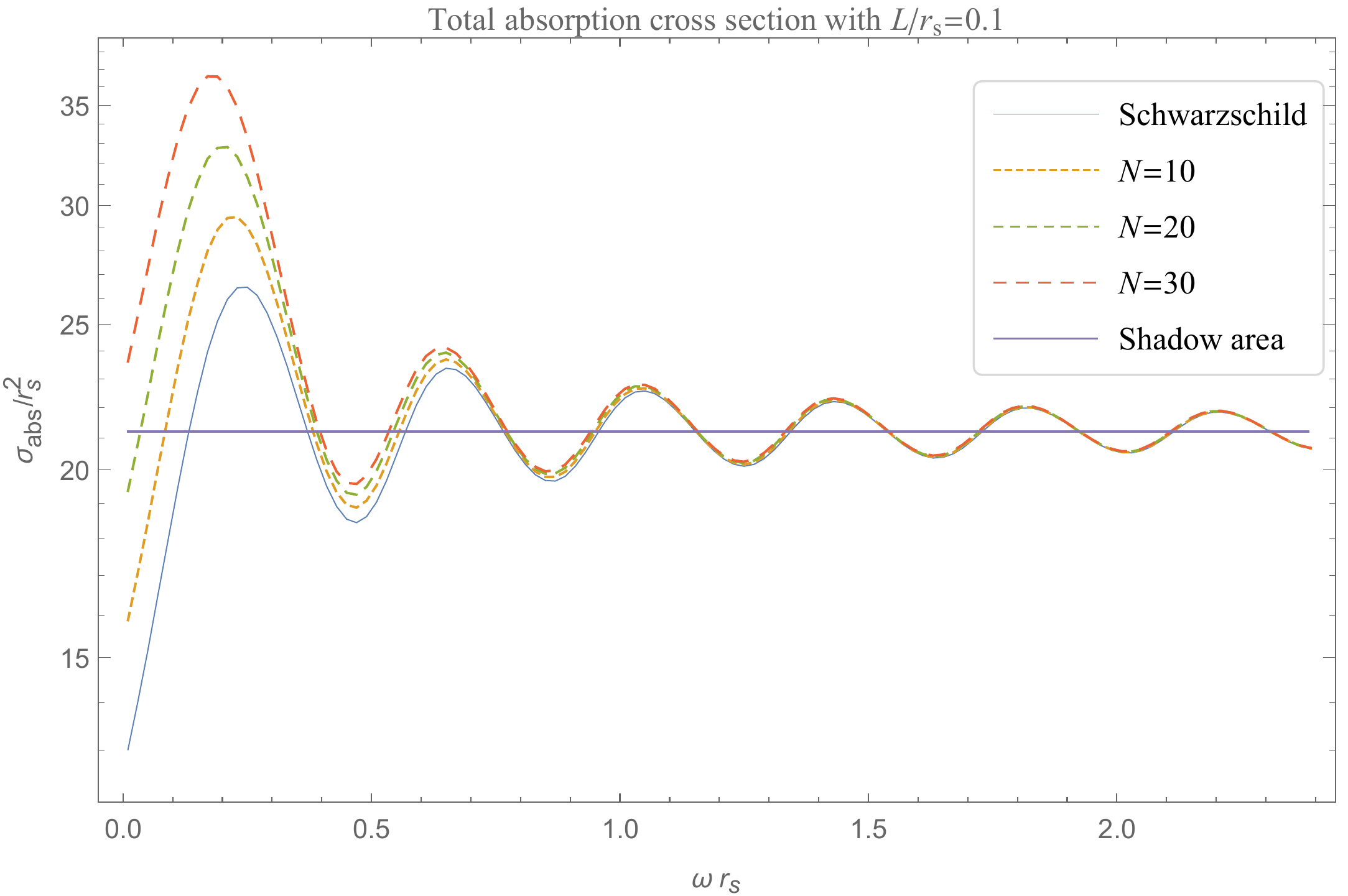}
	\includegraphics[width=0.8\linewidth]{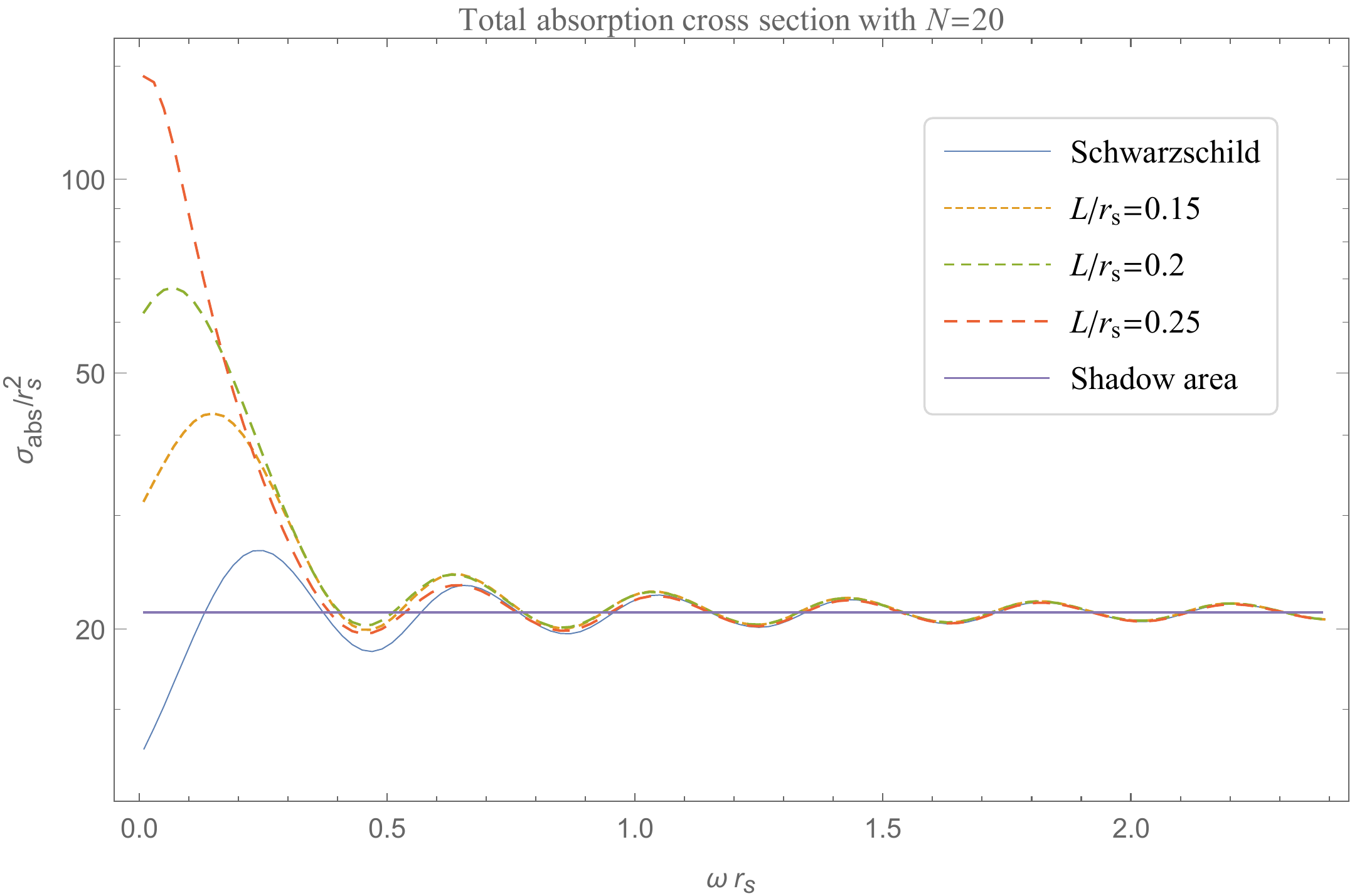}
	\caption{The total absorption cross sections with various $N$ and $L$
	for nonsingular static spherical black holes in conformal gravity.
The straight lines indicate the high frequency limit ($\omega r_\text s\rightarrow\infty$).}
	\label{abstt}
\end{figure}
In Figs.\ref{absl=0} and \ref{abstt} we illustrate the partial ($l=0$) and total absorption cross sections
of the nonsingular static spherical black holes in conformal gravity
for various parameters $L$ and $N$.
One can see that in the low frequency regime where $l=0$ part dominates,
the absorption generally increases with the conformal parameters $L$ and $N$.
This suggests that the net effect of
the potential barrier and well
shown in Fig.\ref{pot}
is generally stronger attraction for larger $L$ or $N$.
Since the effective potential reflects the spacetime geometry,
the dependence of the absorption on the conformal factor
may be related to the conformally rescaled horizon,
namely the ultimate absorbing surface.
In fact, although the horizon still lies at $r=r_\text s$ for any $L$ and $N$,
the area of the horizon is no longer $A_\text{s}=4\pi r^2_\text s$.
Instead, it should be rescaled by a conformal factor
as the spatial slice of the metric is conformally transformed,
i.e., $A=S(r_\text s)A_\text{s}$.
It can be checked that in the $\omega\rightarrow0$ limit,
the absorption cross section tends to the rescaled area $A$ of the horizon\cite{PhysRevLett.78.417,Higuchi:2001si,Avery2017}.
Nonetheless, the crosses of the absorption lines in Figs.\ref{absl=0} and \ref{abstt}
embody the complicated wrestling of the higher potential barrier and deeper potential well seen in Fig.\ref{pot}.
Note that the absorption cross sections are seen far away from the black holes.
The difference in absorption depending on the conformal parameters
indicates that one can judge from afar
whether the central black hole is a Schwarzschild one or a black hole that can be
described by the conformal gravity with a conformal transformation given in Eqs.\eqref{conformaltransf} and \eqref{Sr}.

In the high frequency regime ($\omega r_\text s\gg1$)
where the wave length is much smaller than the size of the horizon,
the absorption cross section should tend to the shadow area of the black hole\cite{PhysRevD.83.044032}.
As mentioned in Sec.\ref{sec:bhs},
in the current cases the shadow area is identical
for any choice of conformal parameters $L$ and $N$.
It can be seen from Fig.\ref{abstt} that
the absorption cross sections in different cases
indeed tend to the same shadow area (the straight line) in high frequency limit,
which is in concordance with the Schwarzschild shadow $A_\text{shadow}$.
Moreover, the oscillation of the absorption cross section with respect to the frequency $\omega$ can be approximated by $\sim\mathrm{sinc}(T_0\omega)$\cite{PhysRevD.83.044032},
where $\mathrm{sinc(x)}$ is the sine cardinal $\sin x/x$,
and $T_0$ is the period for a massless particle to orbit on the photon sphere.
Due to the identical classical null orbits,
one can see from Fig.\ref{abstt} that the different cases demonstrate almost the same oscillation in the high frequency regime.
The similar behaviors to the Schwarzschild case of wave absorption in high frequency regime
help distinguish the nonsingular black holes in conformal gravity from
those spherical black holes which have different high frequency absorption depending on their model parameters\cite{PhysRevD.79.064022,
PhysRevD.92.024012,
liao2012,
deOliveira2018,
Glampedakis_2001}.

\section{Scattering cross section}%
\label{sec:scattering}
As for the scattering of the scalar wave by the black holes in conformal gravity,
the scattering amplitude can be expressed as
\begin{equation}
	g(\theta)=\frac{1}{2i\omega}\sum_{l=0}^{\infty}\left( 2l+1 \right)\left[ \e^{2i\delta_l(\omega)}-1 \right]\mathrm P_l\left( \cos\theta \right),
	\label{scatteramp}
\end{equation}
where
\begin{equation}
	\e^{2i\delta_l(\omega)}\equiv(-1)^{l+1}\mathcal R_{\omega l}
	\label{scatterphaseshift}
\end{equation}
with $\mathcal R_{\omega l}$ given in Eq.\eqref{asymptpsi}
indicates the phase shifts of the scattered waves and $\mathrm P_l$ is the Legendre polynomial.
The differential scattering cross section is given by
\begin{equation}
	\frac{\diff\sigma_\text{sc}}{\diff\Omega}=\left|g(\theta)\right|^2.
	\label{diffscattering}
\end{equation}

Near the antipodal direction $\theta\simeq\pi$,
the scattering interference usually results in a bright spot.
This is known as glory scattering,
which can be approximated analytically by\cite{Matzner1968}
\begin{equation}
	\left.\frac{\diff\sigma_\text{g}}{\diff\Omega}\right|_{\theta\approx\pi}\simeq2\pi\omega b^2_\text g\left|\frac{\diff b}{\diff\theta}\right|_{\theta=\pi}J_0^2\left( \omega b_\text g \sin\theta \right).
	\label{glory}
\end{equation}
Here $b$ is the impact parameter and $b_\text g$ is the one that corresponds to the scattering angle $\theta=\pi$.
$J_0$ is the Bessel function of the first kind with order $0$.

\begin{figure}[!htpb]
	\centering
	\includegraphics[width=0.8\linewidth]{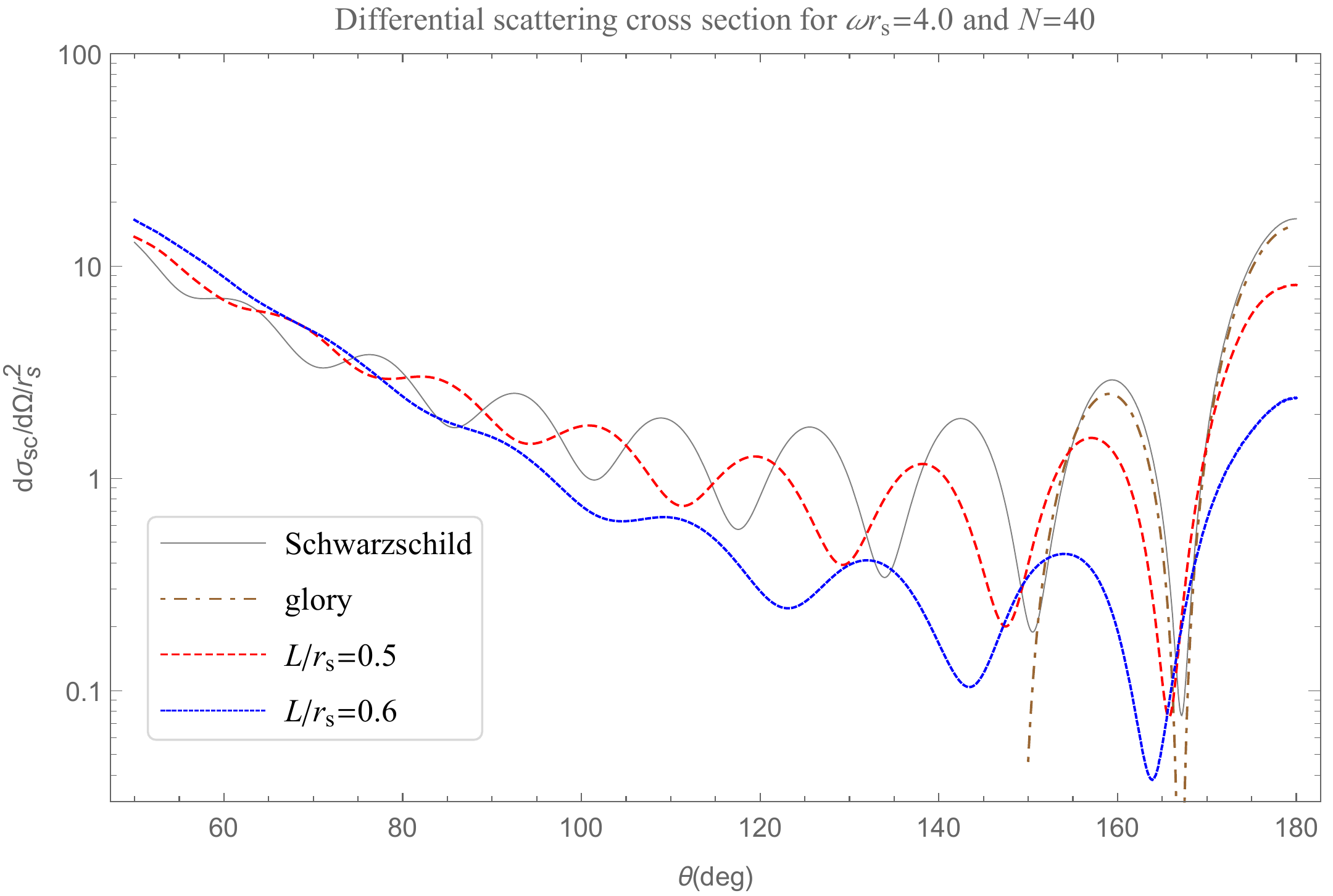}
	\includegraphics[width=0.8\linewidth]{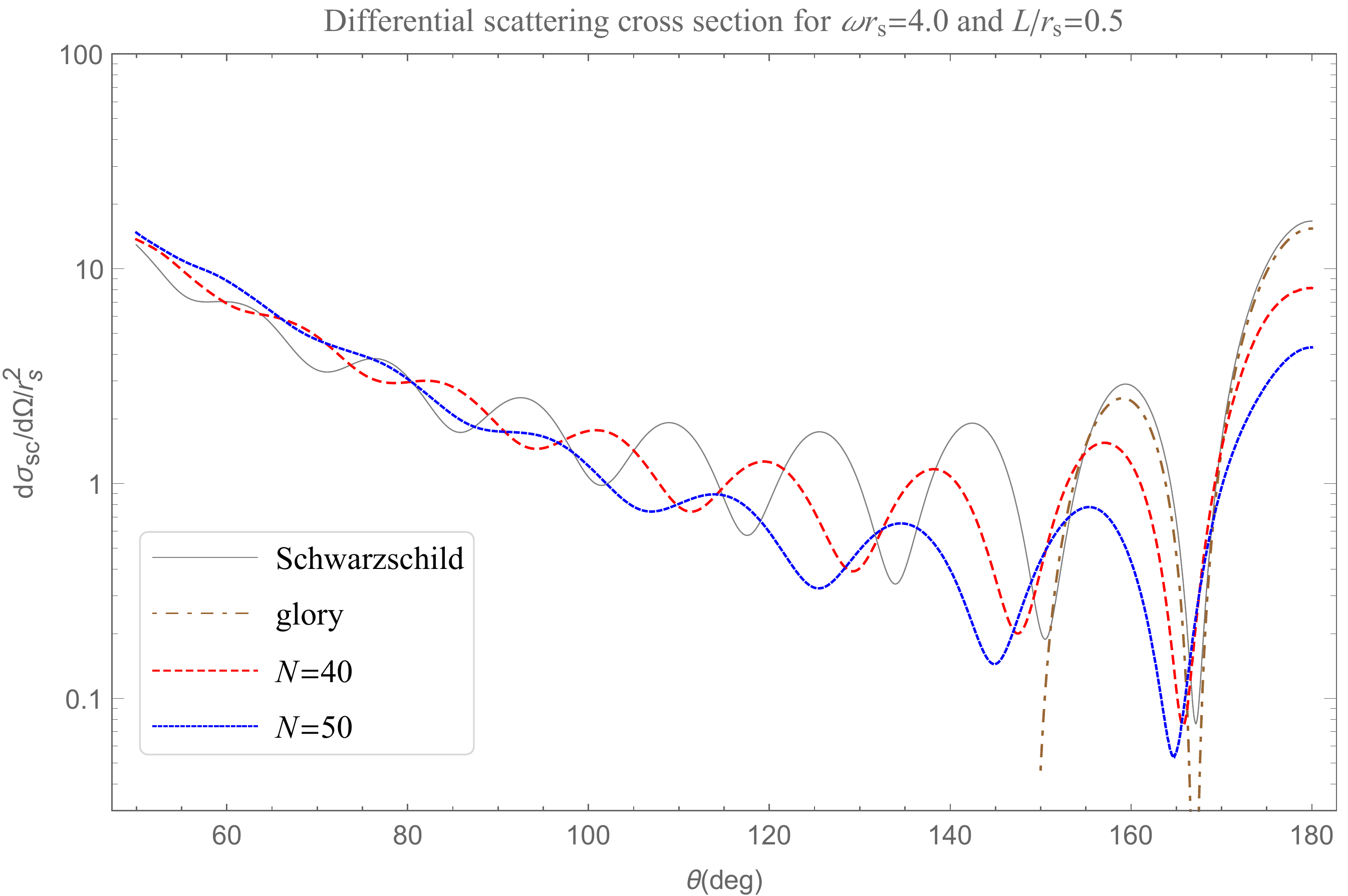}
	\caption{The differential scattering cross sections with $\omega r_\text s=4.0$ and various $N$ and $L$
	for nonsingular static spherical black holes in conformal gravity.
The backward glory scattering is also plotted.}
	\label{sc4}
\end{figure}
In Fig.\ref{sc4} we plot the differential scattering cross sections of the nonsingular static spherical black holes in conformal gravity
using Eq.\eqref{diffscattering}
\footnote{In actual calculation, we have employed the method developed in Ref.\cite{PhysRev.95.500} to improve the convergence.},
as well as the backward glory computed from Eq.\eqref{glory}.
The width of the glory peak is prescribed by $b_g$ in the argument of the Bessel function in Eq.\eqref{glory}.
For some black holes (see, e.g., Refs.
\cite{PhysRevD.79.064022,
PhysRevD.92.024012,
liao2012,
deOliveira2018,
Glampedakis_2001}),
the null orbits, and hence $b_g$, are altered by model parameters.
Consequently, the widths of the glory peaks in these black holes
are significantly different for various model parameters.
However, as mentioned before,
in the current case of nonsingular black holes in conformal gravity,
the geometric optics seems the same as the Schwarzschild spacetime from afar,
and the deflection equation $b(\theta)$ can still be described by Eq.\eqref{darwin}.
It follows that the widths of the glory peaks (see Fig.\ref{sc4})
for various conformal parameters do not differ drastically
like the other aforementioned black holes.

The oscillatory behavior in the scattering cross section is known to come from wave orbiting\cite{PhysRevD.46.4477}.
Roughly speaking,
the wave deflected to observing angle $\theta$ interferes with those deflected to $2n\pi\pm\theta$, $n\in\mathbb N$.
Since the conserved angular momentum is modified by the conformal factor (see Eq.\eqref{conservedel}),
the phases of the waves reaching the observer are therefore also modified (see Appendix for a simplified picture).
This makes the peaks (or troughs) shift towards smaller observing angles
as the conformal parameters $N$ or $L$ grows.
This effect also helps tell apart the actual scattering peaks
and the glory approximation Eq.\eqref{glory} of the Schwarzschild black hole
in the observing angles slightly away from the antipodal direction.
Moreover, the heights of the peaks in the scattering patterns
are lower due to greater absorption (see Fig.\ref{abstt})
for larger $L$ or $N$.

\begin{figure}[!htpb]
	\centering
	\includegraphics[width=0.8\linewidth]{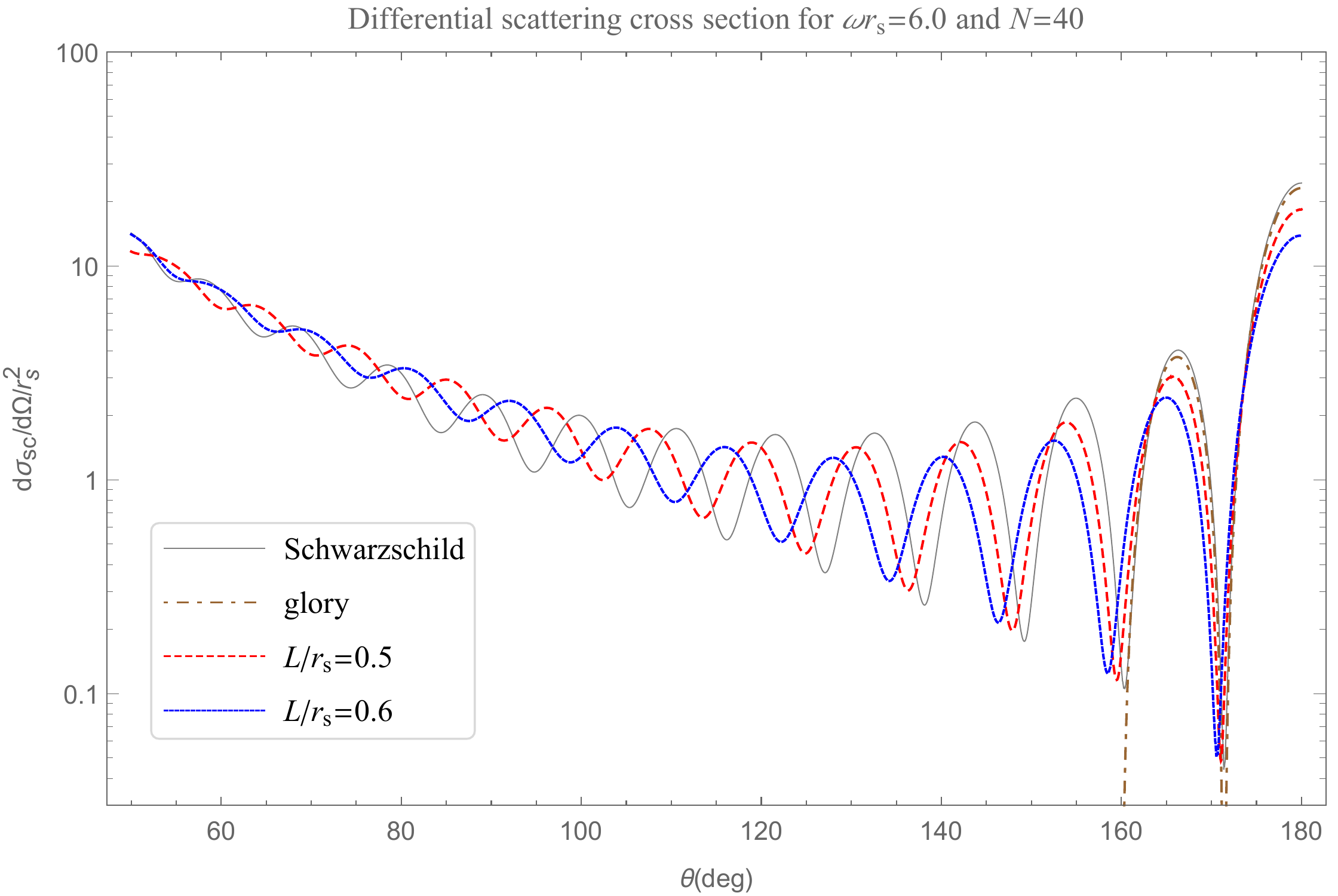}
	\includegraphics[width=0.8\linewidth]{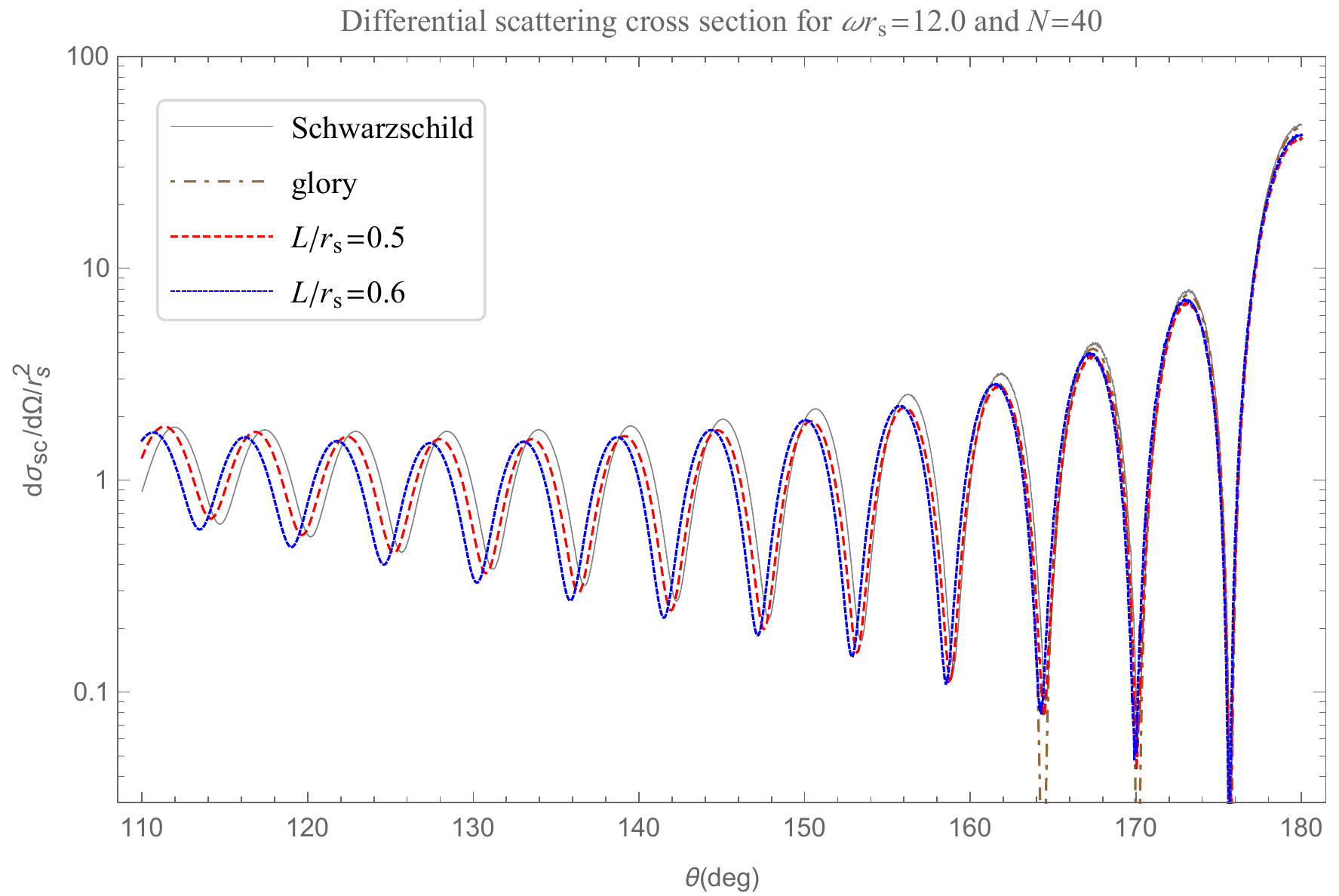}
	\caption{The differential scattering cross sections with $\omega r_\text s=6.0$ and $12.0$
	for nonsingular static spherical black holes in conformal gravity.
	For comparison, we have only plotted the scattering cross sections
	in the observing angle range $110^\circ\sim180^\circ$ for the $\omega r_\text s=12.0$ case.}
	\label{sc6}
\end{figure}

In Fig.\ref{sc6},
we plot the differential scattering cross sections for $\omega r_\text s=6.0$ and $12.0$.
Comparing the two graphs in Fig.\ref{sc6} and the upper panel of Fig.\ref{sc4},
we can see that the difference of the oscillatory behavior
in the scattering cross sections among the cases of different $L$
(Schwarzschild case corresponds to $L/r_\text s=0$.) becomes smaller
as the frequency gets higher.
This is because in the high frequency regime,
as the wave length becomes more negligible,
the wave behaves more classically
and can be better described by the same null orbit for different conformal factors.

\section{Conclusions and discussions}%
\label{summary}
In the present paper,
we have looked into the absorption and scattering of impinging massless scalar waves
by nonsingular static spherical black holes
in order to probe the conformal invariance in the region close to them.
We have shown the absorption and scattering cross sections for typical
choices of conformal parameters $L$ and $N$.

The geometric optics seen afar
is the same as the Schwarzschild spacetime.
Therefore, when the wave length of the impinging wave
is negligible compared to the horizon scale,
it behaves a lot like in the Schwarzschild case.
This gives identical shadow areas of the nonsingular static spherical black holes,
which helps distinguish the conformally invariant spacetime from
those static spherical black holes with altered null orbits\cite{PhysRevD.79.064022,
PhysRevD.92.024012,
liao2012,
deOliveira2018,
Glampedakis_2001}.

Away from the high frequency limit,
where the wave properties are somewhat significant,
larger conformal parameters $L$ or $N$
means greater horizon area $A$.
This geometric effect on interaction between the impinging wave and the spacetime
also incarnates itself in the effective potential,
providing not only a higher barrier but also a deeper well for larger conformal parameter $L$ or $N$.
The overall consequence is that the absorption cross section of the nonsigular static spherical black hole in conformal gravity
generally increases with the conformal parameter $L$ and $N$,
which can be used to tell apart the nonsingular black holes from the Schwarzschild one.
This difference in absorption for various conformal parameters
is in concordance with the different heights of the peaks in the scattering pattern.

Besides the heights, conformal factor also leads to shifts of the oscillatory peaks in the scattering cross section.
In the literature concerning black hole scattering,
the oscillatory pattern shifts are mainly due to the null geodesics altered by,
e.g. the charge of the Reissner-Nordstr\"om black hole\cite{PhysRevD.79.064022}
or the Bardeen black hole\cite{PhysRevD.92.024012},
the coupling in Ho\v rava-Lifshitz gravity\cite{liao2012},
the tidal charge in the brane-world models\cite{deOliveira2018},
or even the angular momentum of the black hole if the wave impinges along the axial direction\cite{Glampedakis_2001}.
Hence in these cases the widths of the glory peaks are obviously different from the Schwarzschild black hole.
However, this is not the case for the nonsingular static spherical black holes in conformal gravity.
The conformal factor does not change the null orbits,
but it alters the phases of the waves that interferes with each other at a given observing angle.
Thus the scattering cross sections of the nonsingular static spherical black holes
also have unique pattern:
the glory peaks, unlike other aforementioned black holes, are not significantly different in width,
and the shifts of peaks are perceivable only at the observing angle away from the antipodal direction.

In summary,
nonsingular static spherical black holes in conformal gravity can be told apart from the Schwarzschild case
through its absorption and scattering of a scalar wave away from the high frequency limit;
and the glory peaks and geometric optics can help distinguish it from
other spherical black holes of alternative gravities that change null orbits.
With the rapidly growing observation and more accurate measurements
of the interaction between black holes and all kind of waves,
it is possible that the scattering and absorption patterns of black holes may be observed one day in the near future.
Our study shows that if the spacetime around a static spherical black hole is described by a conformal theory
and is invariant under the conformal transformation,
such a black hole may demonstrate different signatures from the GR prediction,
it may as well separates itself from other alternative gravitation theories.

The investigation in this work may be extended to more complicated cases.
In conformal gravity, nonsingular rotating black holes
have also been given in the literature
\cite{Bambi_2017,PhysRevD.95.064006,PhysRevD.98.024007},
which may be a more realistic subject to consider absorption and scattering
since black holes in reality are more likely to have spins.
The interaction between waves and the Kerr black hole has been well described in GR\cite{Glampedakis_2001,PhysRevD.88.064033},
and the conformally related case may also have unique characteristics.
Moreover, the interactions between conformally nonsingular black holes
and other kinds of waves, such as eletromagnetic and gravitational waves,
are worth further consideration as well.
We will continue to study these issues in the future.

\section*{Acknowledgement}%
\label{sec:acknowledgement}
The authors would like to thank Professor Chao-Jun Feng for helpful discussions. This work is supported by the National Science Foundation of China under Grants No. 11847080
and No. 10671128
and by the Key Project of Chinese Ministry of Education under Grant No. 211059.

\bibliographystyle{apsrev}
\bibliography{ref}
\appendix
\section{The shifts of the scattering pattern}%
\label{app}
Here we show a simplified and somewhat classical picture of
how conformal factor affects the scattering interference.
A more comprehensive analysis can be found in, e.g., Ref.\cite{PhysRevD.46.4477}.

\begin{figure}[!htpb]
	\centering
	\includegraphics[width=0.8\linewidth]{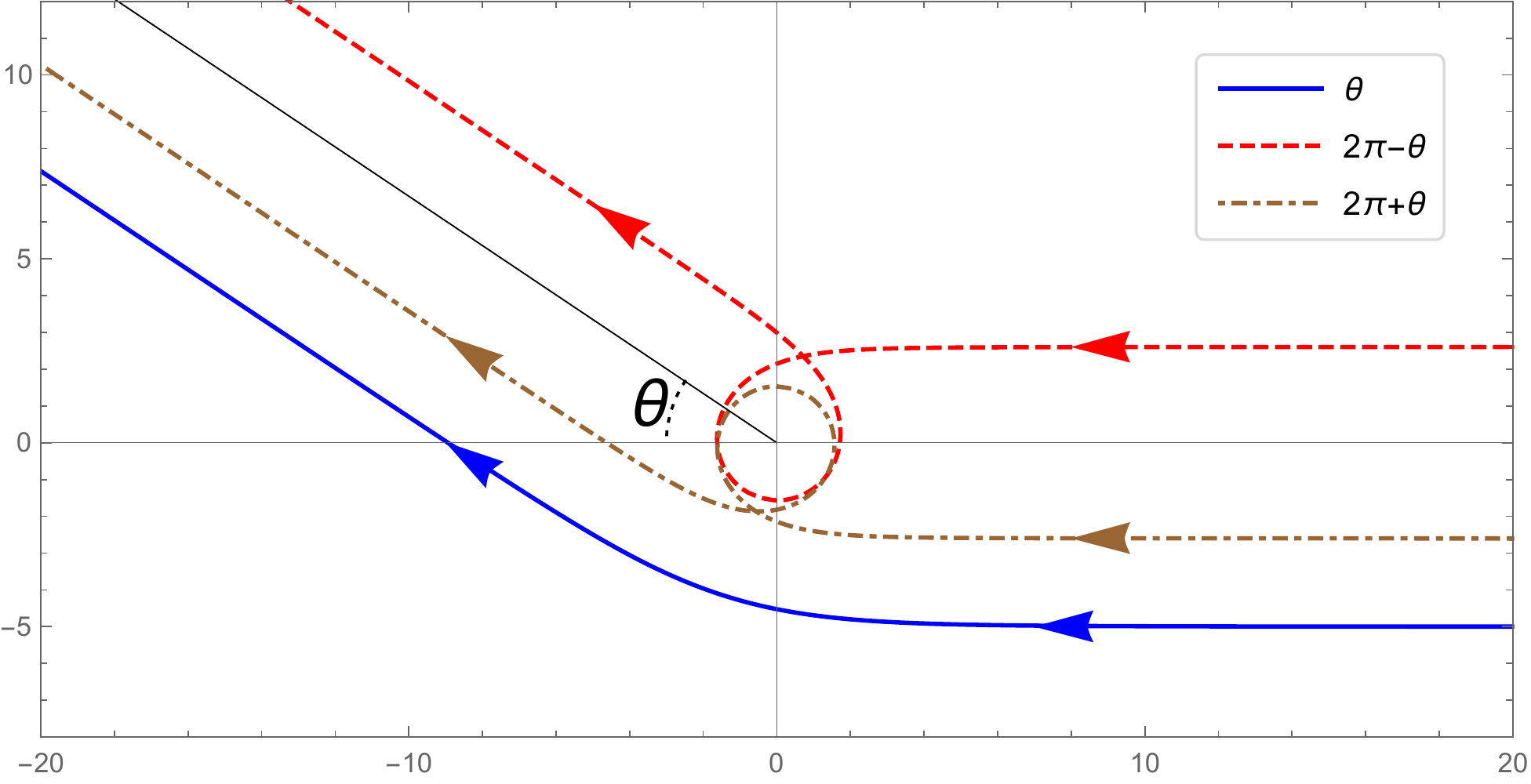}
	\caption{The classical paths of waves deflected to $\theta$, $2\pi-\theta$ and $2\pi+\theta$,
	which will interfere with each other at the observing angle $\theta$.}
	\label{geo}
\end{figure}

Consider the waves deflected to $\theta$ and $2\pi-\theta$ which are interfering with each other,
see Fig.\ref{geo}.
The propagators of the waves are
\begin{equation}
	K\sim\e^{\frac i\hbar\mathcal S},
	\label{propagtors}
\end{equation}
with
\begin{equation}
	\begin{split}
		\mathcal S=&\int \mathcal L\diff t-\vec p_\text f\cdot\vec q_\text f\\
		=&\int\vec p\cdot\diff\vec q-e(t_\text f-t_\text i)-\vec p_\text f\cdot\vec q_\text f,
	\end{split}
	\label{propaction}
\end{equation}
where the integral is taken along the classical path,
and the subscripts i and f indicate the initial and final points of the path, respectively.
In the plane with $\dot\phi=0$,
$\vec p\cdot\diff\vec q=p_r\diff r+p_\theta r\diff\theta$.
Using Eq.\eqref{conservedel}, we have
\begin{equation}
	\int\vec p\cdot\diff\vec q=j\int\frac{\diff\theta}{S(r)}+\int p_r\diff r,
	\label{intp}
\end{equation}
where $j$ is the conserved angular momentum and $S(r)$ is the conformal factor Eq.\eqref{Sr}.
Since $e(t_\text f-t_\text i)$ and $\int p_r\diff r-\vec p_\text f\cdot\vec q_\text f$ do not depend significantly on $\theta$,
the integral over $\theta$ in the phase is the only part that is sensitive to the observing angle.
For the two classical paths of the scattered waves being considered,
\begin{equation}
	K_\theta\sim\mathcal A_\theta\exp\left[\frac i\hbar j_\theta\int_0^{\pi+\theta}\frac{\diff\theta}{S(r)}\right],\quad K_{2\pi-\theta}\sim\mathcal A_{2\pi-\theta}\exp\left[\frac i\hbar j_{2\pi-\theta}\int_0^{3\pi-\theta}\frac{\diff\theta}{S(r)}\right],
	\label{k12}
\end{equation}
where $\mathcal A$ is the complex amplitudes with the angular-insensitive phase absorbed in it.

The differential scattering cross section is
\begin{equation}
	\frac{\diff\sigma}{\diff\Omega}=\frac{|\vec p_\text i|^2}{(2\pi\hbar)^2}\left|K_\theta+K_{2\pi-\theta}\right|^2=\frac{\diff\bar\sigma}{\diff\Omega}\left|1+\frac{K_{2\pi-\theta}}{K_\theta}\right|^2,
	\label{diffra}
\end{equation}
where $\frac{\diff\bar\sigma}{\diff\Omega}$ is the zeroth order approximation and can be taken as, e.g., the Rutherford scattering cross section.
Using Eq.\eqref{k12}, we have
\begin{equation}
	\begin{split}
	\frac{\diff\sigma}{\diff\Omega}=&\frac{\diff\bar\sigma}{\diff\Omega}\left|1+\mathcal A\exp\left[ \frac i\hbar \left( j_{2\pi-\theta}\int_0^{3\pi-\theta}\frac{\diff\theta}{S(r)}-j_\theta\int_0^{\pi+\theta}\frac{\diff\theta}{S(r)} \right) \right]\right|^2\\
	\sim&\frac{\diff\bar\sigma}{\diff\Omega}\left|1+\mathcal A\exp\left[ \frac i\hbar j_\theta\left( \int_0^{3\pi-\theta}\frac{\diff\theta}{S(r)}-\int_0^{\pi+\theta}\frac{\diff\theta}{S(r)} \right) \right]\right|^2\\
	\sim&\frac{\diff\bar\sigma}{\diff\Omega}\left|1+\mathcal A\exp\left[ \frac i\hbar j_\theta \int_{\pi+\theta}^{3\pi-\theta}\frac{\diff\theta}{S(r)} \right]\right|^2.
	\end{split}
	\label{diffra1}
\end{equation}
For nonzero conformal parameter $N$ and $L$, $S(r)>1$, hence
\begin{equation}
	\int_{\pi+\theta}^{3\pi-\theta}\frac{\diff\theta}{S(r)}\equiv\Theta(2\pi-2\theta)<\int_{\pi+\theta}^{3\pi-\theta}\diff\theta=2\pi-2\theta.
	\label{intdiff}
\end{equation}
Obviously $\Theta(2\pi-2\theta)$ is a decreasing function of $\theta$
since smaller $\theta$ means larger upper bound and smaller lower bound of the integral and $S(r)>0$.
Therefore,
if a certain phase, say a peak in the scattering pattern,
is reached at a given $\theta$ in the Schwarzschild case,
the same phase should be reached at a smaller angle $\tilde\theta<\theta$ in the conformal cases
such that $\Theta(2\pi-2\tilde\theta)=2\pi-2\theta$.
This gives a simplified explanation of the shifts of the peaks in the scattering pattern in conformal gravity,
as seen in Fig.\ref{sc4}.

Note that although we have utilized classical paths to simplify the analysis,
in a more comprehensive version of calculation (see, e.g., Ref.\cite{PhysRevD.46.4477}),
the stationary phase approximation would still select the given integral.

\end{document}